\documentclass{PoS}
\usepackage{url}
\usepackage{amsmath,amsbsy,amssymb}
\usepackage{multirow}

\newcommand{\Cyg}{\mbox{Cyg\,X-1}}
\newcommand{\RXTE}{\textsl{RXTE}}

\newcommand{\vskipAfterSection}{\vskip -0.25cm}

\title{Catching Up on State Transitions in Cygnus\,X-1}
\ShortTitle{Catching Up on State Transitions in Cygnus\,X-1}

\author{\speaker{{\large Moritz B\"ock}},\quad\quad
        Manfred Hanke,\quad\quad
        J\"orn Wilms,\quad\quad
        Stefan Pirner\\
        Dr.~Karl Remeis-Observatory, Bamberg\\
        Erlangen Centre for Astroparticle Physics, Germany\\
        E-mail: \email{Moritz.Boeck@sternwarte.uni-erlangen.de}}
\author{Victoria Grinberg,\quad\quad Sera Markoff\\
        Astronomical Insitute ``Anton Pannekoek'', University of Amsterdam, Netherlands}
\author{Katja Pottschmidt\\
        CRESST and NASA Goddard Space Flight Center\\
        Center for Space Science and Technology, University of Maryland
        Baltimore County}
\author{Michael A.~Nowak\\
        Massachusetts Institute of Technology, Kavli Institute for Astrophysics and Space Research, CXC}
\author{Guy Pooley\\
        Cavendish Laboratory, Cambridge, UK}

\abstract{
 In 2005 February we observed Cygnus X-1 over a period of 10 days quasi-continuously
 with the \textsl{Rossi X-ray Timing Explorer} and the Ryle telescope.
 We present the results of the spectral and timing analysis on a timescale of 90\,min
 and show that the behavior of \Cyg{} is similar
 to that found during our years long monitoring campaign.
 As a highlight we present evidence for a full transition from the hard to the soft state
 that happened during less than three hours.
 The observation provided a more complete picture of a state transition than before,
 especially concerning the evolution of the time lags, due to unique
 transition coverage and analysis with high time resolution.

}

\FullConference{VII Microquasar Workshop: Microquasars and Beyond\\September 1-5 2008\\Fo\c{c}a, Izmir, Turkey}

\begin{document}

\section{Introduction} \vskipAfterSection
The microquasar Cygnus\,X-1 is usually observed in the low/hard state
or in the high/soft (steep power-law) state -- or in an intermediate state \cite{Wilms2006}.

In the low/hard state the 4--250\,keV \RXTE{} broadband spectrum can be described
by a broken power-law (with $\Gamma\equiv\Gamma_1\lesssim 2.1$\footnote{~
 A definition of the typical hard state spectrum is given by $\Gamma_1\le2.1$ \cite{Wilms2006}.
 Note that the transition from the intermediate hard to the intermediate soft state, which is analyzed in this work, occurs at $\Gamma_1\approx2.4$.
}
for $E<E_\mathrm{break}\approx$10\,keV
and $\Gamma\equiv\Gamma_2<\Gamma_1$ for~$E>E_\mathrm{break}$)
with high-energy cutoff $\propto\exp(-E/E_\mathrm{fold})$ 
(with $120\,{\rm keV}\lesssim E_\mathrm{fold}\lesssim250\,$keV) \cite{Wilms2006},
which is canonically interpreted as the result of Comptonization by thermal electrons.
The variability is large ($\approx30\,\%$ root mean square (rms) amplitude)
due to several broad noise components in the power spectrum (PSD)
between 0.01 and 100\,Hz, which can be modeled with Lorentzian profiles \cite{Nowak1999timing,Pottschmidt2003}.
In the hard state \Cyg{} emits a flat radio spectrum with a flux of $\sim$15\,mJy \cite{Fender2000}.

In~the soft state the soft X-ray spectrum is dominated by thermal emission from the accretion disk,
which is believed to extend close to the innermost stable orbit around the black hole.
The power-law component of the spectrum is steeper,
but often extends into the $\gamma$-ray regime without notable cutoff,
indicating that Comptonization by non-thermal electrons
is much more important in the soft state than in the hard state
\cite[see also the contribution of J.~Malzac to this volume]{McConnell2002,CadolleBel2006}.
The radio emission in the soft state is quenched.
The timing properties are also very different:
the PSD is a $f^{-1}$ powerlaw  with cutoff at $\approx10\,$Hz,
without the broad noise components and thus a much lower variability \cite{Pottschmidt2003}.

In the intermediate state X-ray and radio flares can be observed \cite{Wilms2007}. \Cyg{} often shows failed state transitions instead of full transitions \cite{Pottschmidt2000}. To determine the nature of the transitions it is helpful to combine spectral and timing properties. The peak frequencies of the Lorentzians in the power spectra are correlated with the photon index \cite{Pottschmidt2003,Axelsson2005}. Time lags between light curves in different energy bands depend on the photon index as well, the correlations can be explained with models based on Comptonization in a jet \cite{Kylafis2008}. How some of these correlations change after the transition is discussed in this work.

\section{Observation and Data Analysis} \vskipAfterSection
\begin{figure}\centering
 \begin{minipage}{0.48\textwidth}
  \includegraphics[width=\columnwidth]{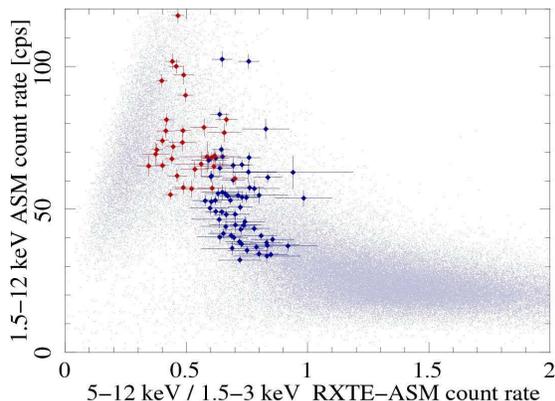}
 \end{minipage}
 \hfill
 \begin{minipage}{0.48\textwidth}
  \caption[]{
   Hardness-intensity-diagram of \Cyg{} from \RXTE-ASM.
   The gray dots show the long term behaviour of \Cyg{}
   from 1996~January until 2008~July by single 90\,min~dwells.
   The hard (right) and soft (left) states are clearly separated.
   The density of dots measures how much time the source spends in each state.
   The red (soft-intermediate) and blue (hard-intermediate) data points
   were obtained within a 10 day period and show the transition in 2005 February,
   which is the topic of this paper.
  }
  \label{fig:asm}
 \end{minipage}
\end{figure}

\begin{figure}\centering
 \begin{minipage}{0.48\textwidth}
  \includegraphics[width=\columnwidth]{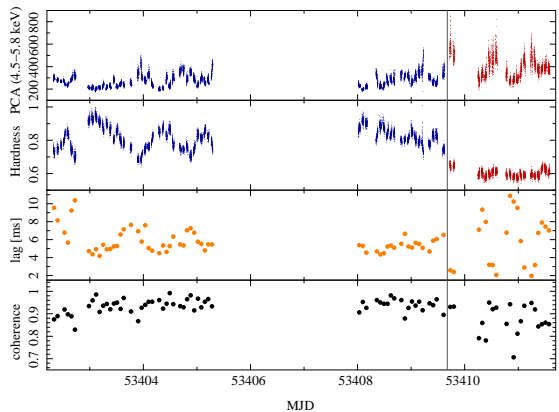}
 \end{minipage}
 \hfill
 \begin{minipage}{0.48\textwidth}
  \caption[]{
   Temporal behavior of the count rate, the hardness, the time lag, and the coherence. The latter two were computed between the
   9.5--15\,keV and 4.5--5.8\,keV light curves.
   The vertical line indicates the transition from the hard-intermediate to the soft-intermediate state.
   A positive time lag means that the high energy light curve has a delay
   with respect to the one in the low energy band.
  }
  \label{fig:multi}
 \end{minipage}
\end{figure}

In 2005, we observed \Cyg{} quasi-continously
with the \textsl{Rossi X-ray Timing Explorer} (\RXTE)
from Feb~1,~10:48 to Feb~4,~7:11~UTC (MJD 53402.45--53405.3)
and from Feb~7,~0:21 to Feb~10, 14:14~UTC (MJD 53408.01--53411.59),
and the Ryle telescope.
The detailed results of our analysis will be described by B\"ock et al.~(2009)~\cite{Boeck2009}.
A hardness-intensity diagram from \RXTE-ASM data (Fig.~\ref{fig:asm})
already shows that \Cyg{} was in an intermediate state during this time.
Because of the fast and strong variability due to flaring,
we split the data into single \RXTE{} orbits of $\sim$3\,ks exposure.

\begin{figure}\centering
 \begin{minipage}{0.48\textwidth}
  \includegraphics[width=\columnwidth]{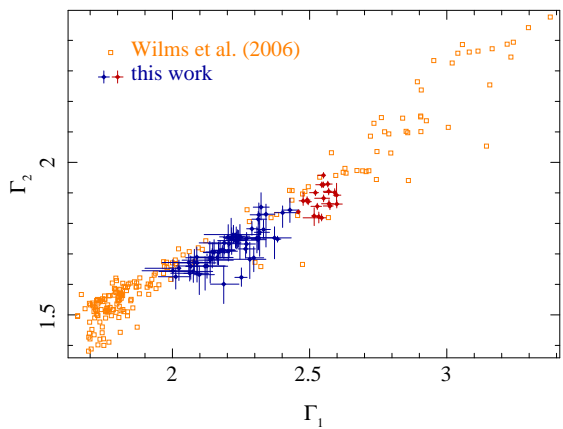}
 \end{minipage}
 \hfill
 \begin{minipage}{0.48\textwidth}
  \caption[]{
   Relation between the photon indices. The linear correlation found in this work is identical to that found
   in the long term spectral evolution from 1999--2004 \cite{Wilms2006}.
   This correlation holds throughout all states. The transition occured at $\Gamma_1=2.4$.
  }
  \label{fig:G}
 \end{minipage}
\end{figure}
For each of these orbits, we extracted the spectrum
and PCA light curves in the 4.5--5.8\,keV
and 9.5--15\,keV energy bands. Using these light curves PSDs were computed, which we call $\rm PSD_{lo}$ and $\rm PSD_{hi}$ in the following.
Between these light curves we calculated coherence and time lags \cite{Nowak1999timing}. Consistent with previous observations the 9.5-15 keV light curve lags the one at lower energies \cite{Nowak1999timing, Pottschmidt2003}.

The spectra were fitted with absorbed broken power-law models with exponential cutoff.
During this observation, the photon index varied in the range $2\le\Gamma_1\le2.6$,
which is more than a third of the range $1.65\le\Gamma_1\le3.4$
found for the long term evolution of \Cyg{} from 1999 to 2004 \cite{Wilms2006}.
The same relation between the two photon indices was found (Fig.~\ref{fig:G}).

We fitted the power spectra with two broad Lorentzian profiles (L$_1$ and L$_2$) 
and an additional power-law with exponential cutoff where required, 
similarly to \cite{Axelsson2005}.

\section{Correlation of the spectral and timing properties} \vskipAfterSection
\begin{figure}\centering
 \begin{minipage}{0.48\textwidth}
  \includegraphics[width=\columnwidth]{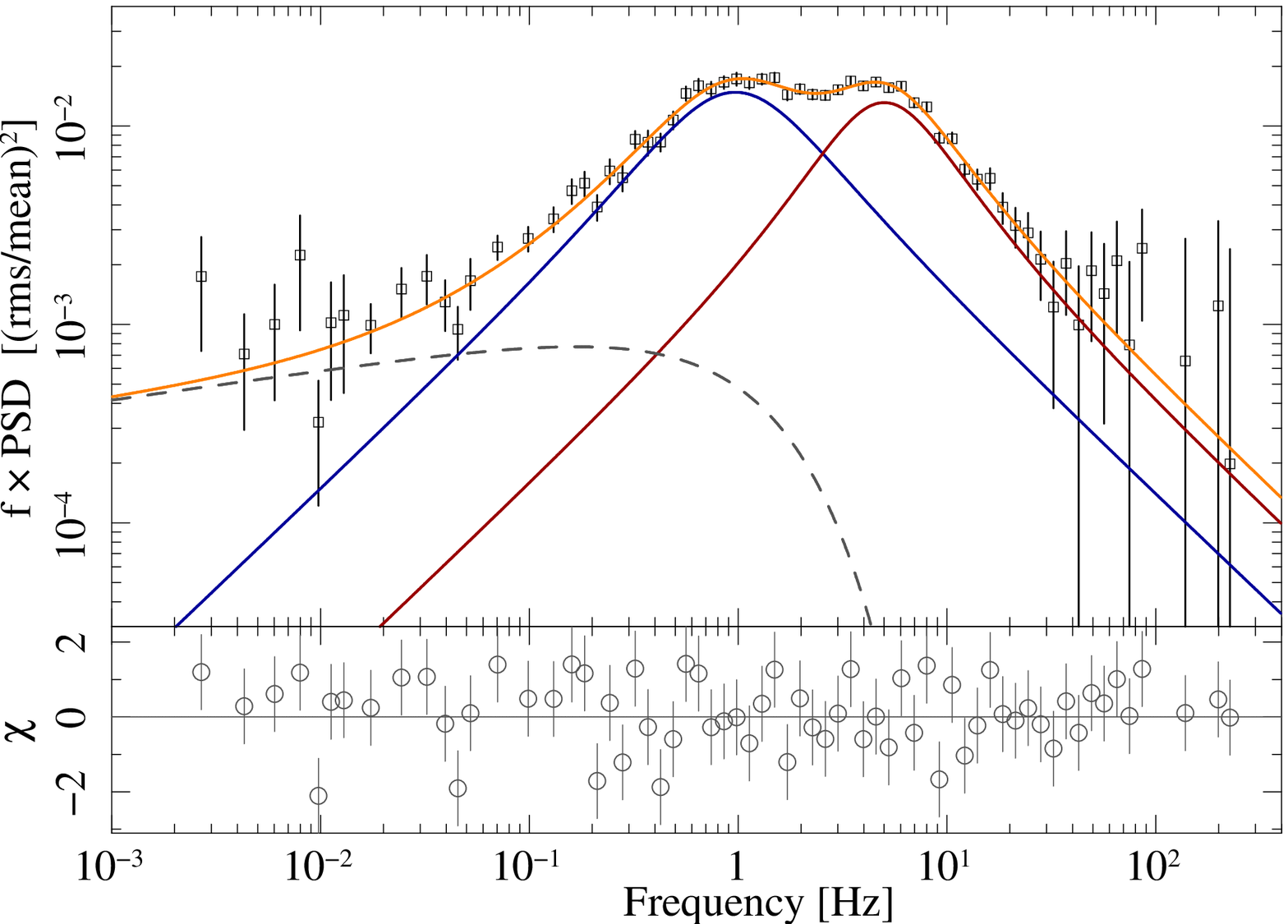}
 \end{minipage}
 \hfill
 \begin{minipage}{0.48\textwidth}
  \includegraphics[width=\columnwidth]{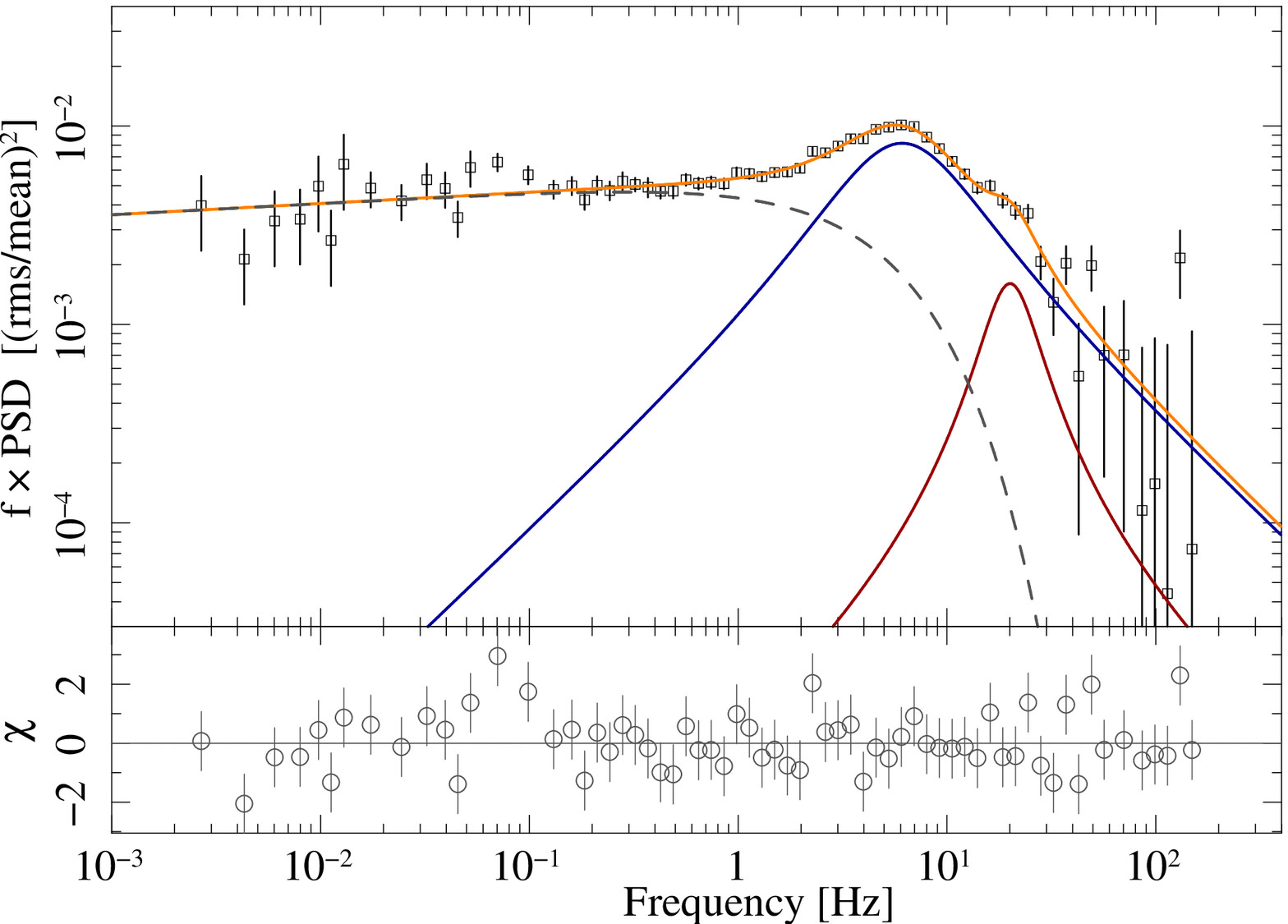}
 \end{minipage}
 \caption[]{
  A PSD$_\mathrm{lo}$ at $\Gamma_1=2.14$ (left) and a PSD$_\mathrm{lo}$ at $\Gamma_1=2.55$ (right),
  modeled as sum of two Lorentzian profiles (L$_1$: blue curve, L$_2$: red curve),
  and a power-law with exponential cutoff (gray dashed line).
 }
 \label{fig:psd}
\end{figure}

\begin{figure}\centering
 \begin{minipage}{0.48\textwidth}
  \includegraphics[width=\columnwidth]{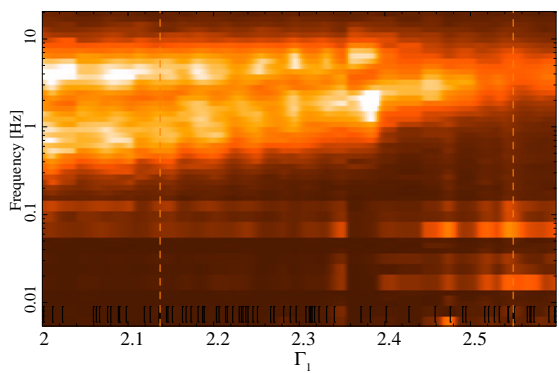}
 \end{minipage}
 \hfill
 \begin{minipage}{0.48\textwidth}
  \caption[]{
   PSD$_\mathrm{lo}$ as a function of $\Gamma_1$:
   For each observation (black dashes at the bottom),
   the values of $f\times\text{PSD}_{\text{lo}}(f)$
   are color-coded (brighter for larger values)
   and interpolated between the corresponding photon indices.
   The vertical dashed lines mark the PSDs shown in Fig.~\ref{fig:psd}.
   The two bright streaks between $\sim$0.5 and $\sim$10\,Hz 
   are the peaks of L$_1$ and L$_2$ in the PSDs,
   which obviously shift to higher frequencies with increasing $\Gamma_1$.
   After the transition at $\Gamma_1\approx2.4$, L$_2$ fades
   and a ${\rm PSD}\propto f^{-1}$ power-law component emerges at low frequencies.
  }
  \label{fig:psdland}
 \end{minipage}
\end{figure}

\begin{figure}\centering
 \begin{minipage}{0.48\textwidth}
  \includegraphics[width=\columnwidth]{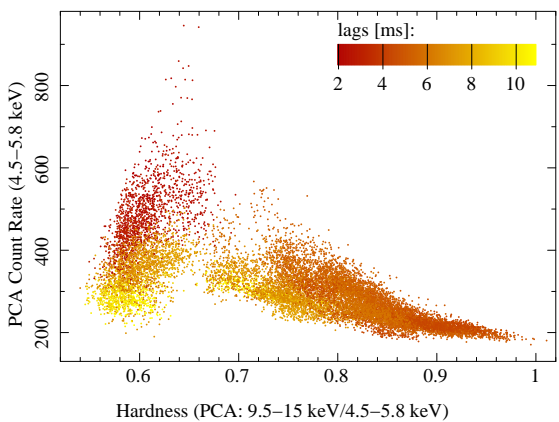}
 \end{minipage}
 \hfill
 \begin{minipage}{0.48\textwidth}
  \caption[]{
   The hardness intensity diagram reveals a state transition.
   After the transition (see Fig.~\ref{fig:multi}),
   the values cover a region in the HID separated from that with larger hardness before.
   The HID is color-coded with the values of the time lags.
   In the soft region we found an interesting relation between the time lags and the count rate.
   A small time lag at high count rates and a large one at low count rates is observed.
   It suggests to subdivide this region into different states.
  }
  \label{fig:hid}
 \end{minipage}
\end{figure}

In comparison with a PSD from a spectrally harder state (Fig.~\ref{fig:psd}, left),
L$_1$ and L$_2$ are shifted to higher frequencies
in a PSD corresponding to a spectrally softer state (Fig.~\ref{fig:psd}, right).
Furthermore the normalization of the Lorentzians, especially that of L$_2$,
-- and therefore the total variability of the source -- decreases significantly,
while the contribution of the power-law increases.
This effect is obvious from the color-coded representation of the PSDs
as a function of the photon index $\Gamma_1$ (Fig.~\ref{fig:psdland}).
A quantitative analysis of the peak frequencies $\nu_i$ and fractional rms contribution rms$_i$
of the Lorentzians $i=1,2$ proves these correlations.
It is interesting that $\nu_2$ is roughly proportional to $\sqrt\nu_1$,
especially when only the data from the hard-intermediate state are used.

A comparison of PSDs in the different energy bands reveales that the normalization of $L_1$ is significantly smaller in PSD$_\mathrm{hi}$ than in PSD$_\mathrm{lo}$, whereas $L_2$ is identical in both energy bands.

An interesting behavior of the time lags was found. In the hard-intermediate state they increase with decreasing hardness of the spectrum, as it was found in \cite{Pottschmidt2003} and modeled in \cite{Kylafis2008}. After the transition to the soft state there is a different behavior. The time lags change with the count rate. At high count rates there is a very low time lag ($\sim$2\,ms) and the light curves in the low and high energy band are strongly coherent, whereas it is the other way around at low count rates (Fig.~\ref{fig:multi} and Fig.~\ref{fig:hid}).

\acknowledgments \vskipAfterSection
We thank the organizing committees for this wonderful conference in Fo\c{c}a\,/\,Izmir, Turkey!
This~work was partially funded by the \textsl{DLR} (Deutsches Zentrum f\"ur Luft- und Raumfahrt) under contract 50OR0701.

\end{document}